\documentclass{PoS}
\usepackage{amsmath}

\title{The KLOE-2 Experiment at DA$\Phi$NE}

\ShortTitle{KLOE-2}

\author{\speaker{Wojciech Wi\'slicki}\thanks{On behalf of the KLOE-2 Collaboration.}\\
        National Center for Nuclear Research\\
        Ho\.za 69, Pl-00-681 Warszawa\\
        E-mail: \email{wojciech.wislicki@ncbj.gov.pl}}

\abstract{We present recent results obtained by the KLOE-2 Collaboration at the DA$\Phi$NE e$^+$e$^-$ collider. The first class of results concerns search for dark forces at the scale of 1 GeV in associated production of $\gamma$ and the U boson, in search for the Higgsstrahlung and in possible decays of $\phi$ into $\eta$ and U.
The second is in neutral kaon physics, on testing the CPT and Lorentz invariance, and on search for quantum decoherence effects in entangled pairs of kaons.
The third class of results concerns precision measurements in hadronic physics at low energy where transition form factors of $\phi$ to pseudoscalar mesons $\pi^0$ and $\eta$ are determined.}

\FullConference{XIII International Conference on Heavy Quarks and Leptons\\
		22-27 May, 2016\\
		Blacksburg, Virginia, USA}

\begin{document}

\section{The KLOE-2 Experiment}
The KLOE-2 Collaboration performs new experiment on the electron-positron collider DA$\Phi$NE at the Laboratori Nazionali di Frascati.
The collider operates at the center-of-mass energy equal to the $\phi(1019)$ mass, or in the vicinity of this energy. 
The KLOE-2 succeeds the KLOE Collaboration \cite{kloe} and provides significant improvements in data quality and precision, detector acceptance, data storage volume and processing speed.

\begin{figure}[h]
\begin{tabular}{ccc}
\includegraphics[scale=.22]{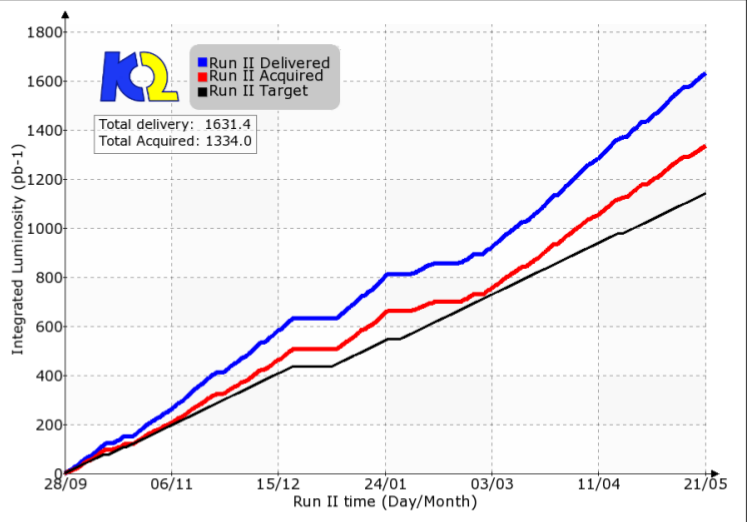} & \includegraphics[scale=.23]{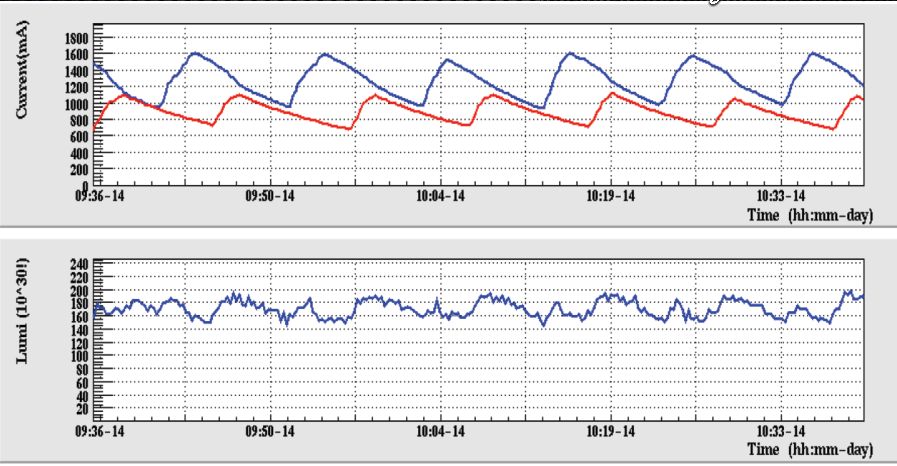} & \includegraphics[scale=.19]{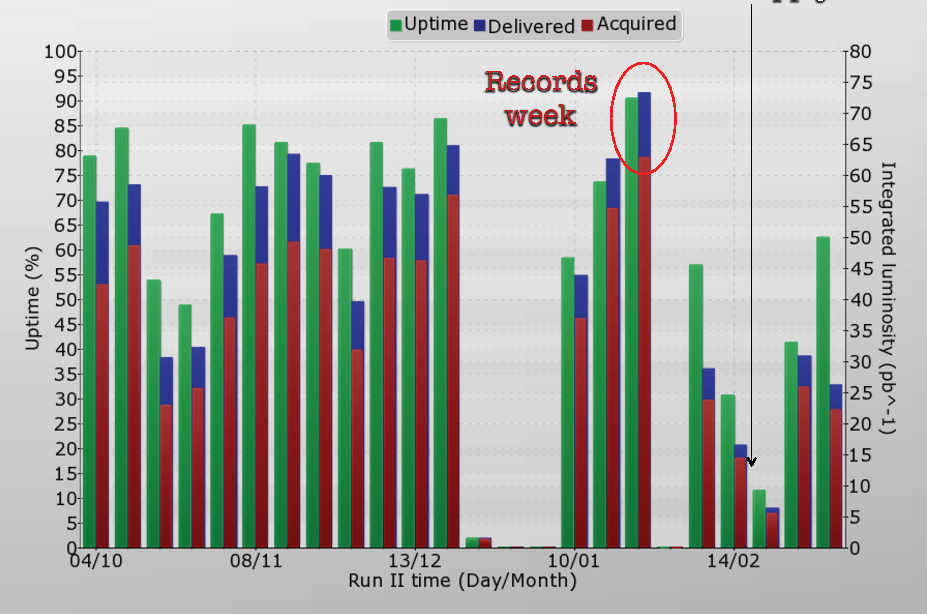}
\end{tabular}
\caption{Illustration of the DA$\Phi$NE and KLOE-2 performance in current data-taking campaign. Left panel: cumulative luminosities from the beginning of the run until the conference's date, where the delivered luminosity is given in blue, the aquired one in red and the target for this run in black. Central panel presents the positron (red) and electron (blue) currents in mA (top) and the instantaneous luminosity (bottom). The rightmost panel present the week-by-week uptime (green) and the delivered (blue) and aquired (red) luminosities.}
\label{fig1}
\end{figure}

Improvement of DA$\Phi$NE luminosity is realized by increasing the Piwinski angle and implementing a new beam optics \cite{crabwaist}. 
For beams of transverse size $\sigma_x$ at their crossing point and colliding at
angle $\theta$, the Piwinski angle depends on these quantities like $\alpha\sim\theta/\sigma_x$. 
In order to make $\alpha$ larger, and thus shrinking beam overlap region, one increases $\theta$ and reduces $\sigma_x$. 
Necessary compensation of betatron resonances is done by using the crab-waist optics realized by two sextupole magnets at both sides of the interaction point. 
In the ongoing data-taking campaign KLOE-2 is aiming to collect at least 5 fb$^{-1}$ in a couple of years. 
At the time of this conference, delivered data volume amounts to 1.7 fb$^{-1}$, as seen in Fig.~\ref{fig1} (left).
In Fig.~\ref{fig1} (center) the instantaneous luminosity and time structure of the positron and electron currents in DA$\Phi$NE are presented, whereas Fig.~\ref{fig1} (right) shows the maximum luminosity delivered weekly, amounting to 76.3 pb$^{-1}$.

Main parts of the spectrometer, originally built for KLOE, are the large-size, cylindrical drift chamber and the electromagnetic calorimeter.
Both detectors cover 98\% of the full solid angle and surround symmetrically the e$^+$e$^-$ interaction point.
The drift chamber, filled in proportions 9:1 with helium and isobutane, and the spectrometer magnet of 0.52 T magnetic field, provide perpendicular vertex resolution of the order of 1 mm and transverse momentum accuracy better than 0.4\%.
The calorimeter, built from lead and scintillation fibers, provides energy resolution $\sigma_E/E=5.7\%/\sqrt{E\mbox{(GeV)}}$ and time resolution $\sigma_t=55/\sqrt{E\mbox{(GeV)}}\oplus 100$ ps.

Detector upgrades and physics program of KLOE-2 are presented in Ref. \cite{kloe2general}. New equipments include:

- The cylindrical inner tracker, consisting of 4 layers of gaseous electron microdetectors (GEM) surrounding interaction point. It improves vertex resolution four times and significantly enlarges acceptance for tracks of low-$p_T$ particles.

- New electron taggers for the high energy ($E>400$ MeV) and low energy ($160<E<230$ MeV) particles, improving acceptance of electrons and positrons scattered at low angles and used in $\gamma\gamma$ physics.

- Calorimeters near the beam: QCALT improving acceptance of the K$_L$ decay products and CCALT improving angular acceptance of photons down to 10$^\circ$.

- Upgrades of the data acquisition system, consisting of new Power-7 computing boards, tape library upgradable to 175 PB, new integrated Power-8 data servers and more efficient network.

In this article we present new results and prospects in three subject areas: search for dark force at the scale of 1~GeV; results on the CP, CPT and Lorentz invariance using neutral kaons; and new precision measurements in hadronic physics at low energy.

\section{Search for Dark Force}
As proposed in Ref. \cite{holdom}, extending the Standard Model by an additional U(1) gauge group entails existence of an additional, spin-1 gauge boson U, also called the dark photon, able to couple to weakly interacting massive candidates for dark matter.
Predictions for rates of their production in resonance decays and e$^+$e$^-$ interactions were given in Ref. \cite{fayet}.

The dark photon can interact via mixing term with ordinary photon and subsequently decay into ordinary particles. 
This can be represented by a mixing term in the Lagrangian
\begin{eqnarray}
{\mathcal L}_{\text{mix}}=\frac{\varepsilon}{2}F_{\mu\nu}G^{\mu\nu},\quad\quad\quad \varepsilon=\alpha_{\text{em}}/\alpha_{\text{dark}},
\label{eq1}
\end{eqnarray}
where $F_{\mu\nu}$ stands for the electromagnetic field, $G_{\mu\nu}$ for dark field, and $\varepsilon$ being the ratio of the electromagnetic to dark coupling constants. The $\varepsilon$ is expected to be in the range $10^{-8}-10^{-2}$.
\begin{figure}[h]
\begin{tabular}{cc}
\hspace{1cm} \includegraphics[scale=.24]{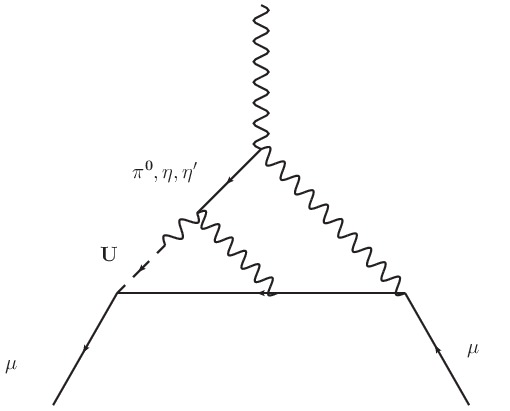} & \hspace{1cm} \includegraphics[scale=.24]{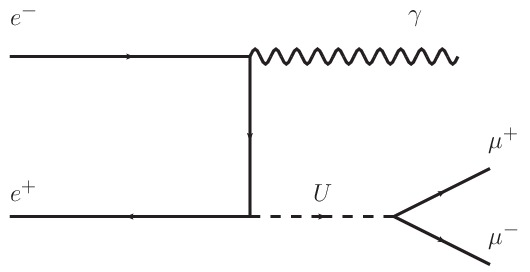} \\
\hspace{1cm} \includegraphics[scale=.24]{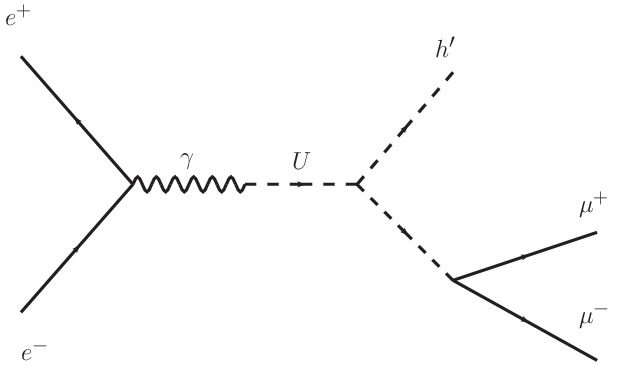} & \hspace{1cm} \includegraphics[scale=.24]{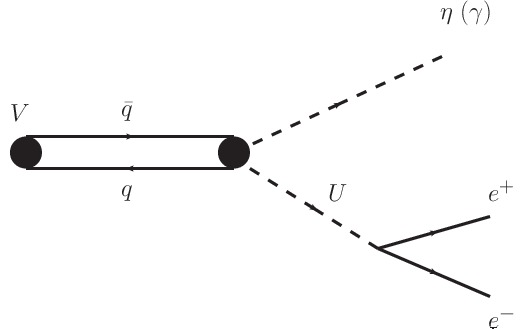}
\end{tabular}
\caption{Contribution of the $\mbox{U}\rightarrow\mu^+\mu^-$ decay to the anomalous magnetic moment of muon (top left); associated production of U and $\gamma$ (top right); the Higgsstrahlung (bottom left); decay $\phi\rightarrow\mbox{U}\eta$ (bottom right).}
\label{fig2}
\end{figure}
For the process of an associated production of U and $\gamma$, Fig. \ref{fig2} (top right), expected signature of U is a narrow peak in mass of the pair of oppositely-charged particles.
KLOE-2 has performed an extensive search for the final-state pairs $\mu^+\mu^-$ \cite{u1}, e$^+$e$^-$ \cite{u2} and $\pi^+\pi^-$ \cite{u3} and experimentally determined 90\% confidence regions in the $\varepsilon$ {\it vs.} M$_{\text{U}}$ plane.  
Possible identification of the process $\mbox{U}\rightarrow\mu^+\mu^-$ is also important for its possible contribution to the g-2 anomaly of $\mu$ (cf. Fig.~\ref{fig2} (top left) and Ref. \cite{hye}).

\begin{figure}[h]
\begin{tabular}{ccc}
\includegraphics[scale=.6]{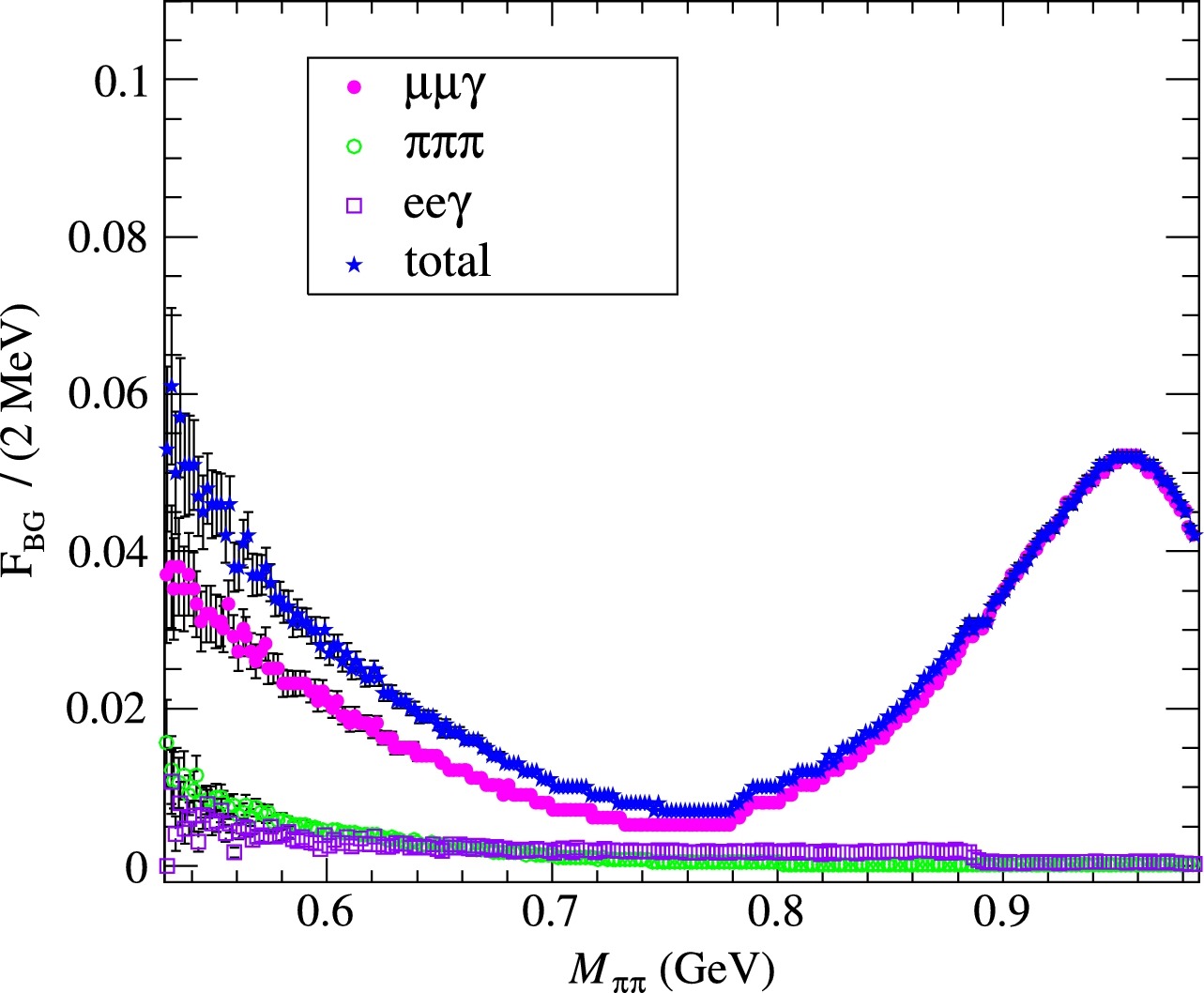} & \includegraphics[scale=.6]{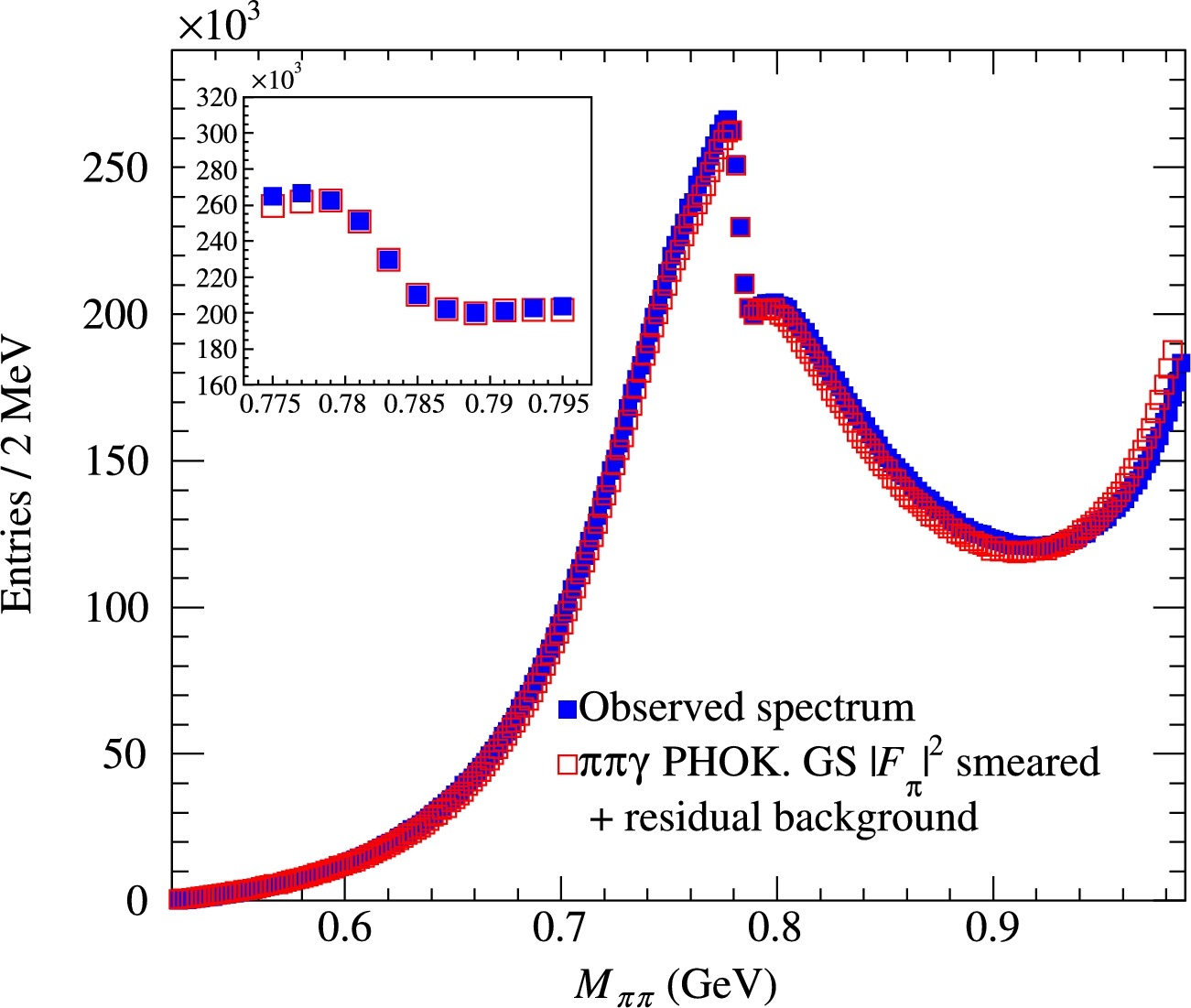} & \includegraphics[scale=.9]{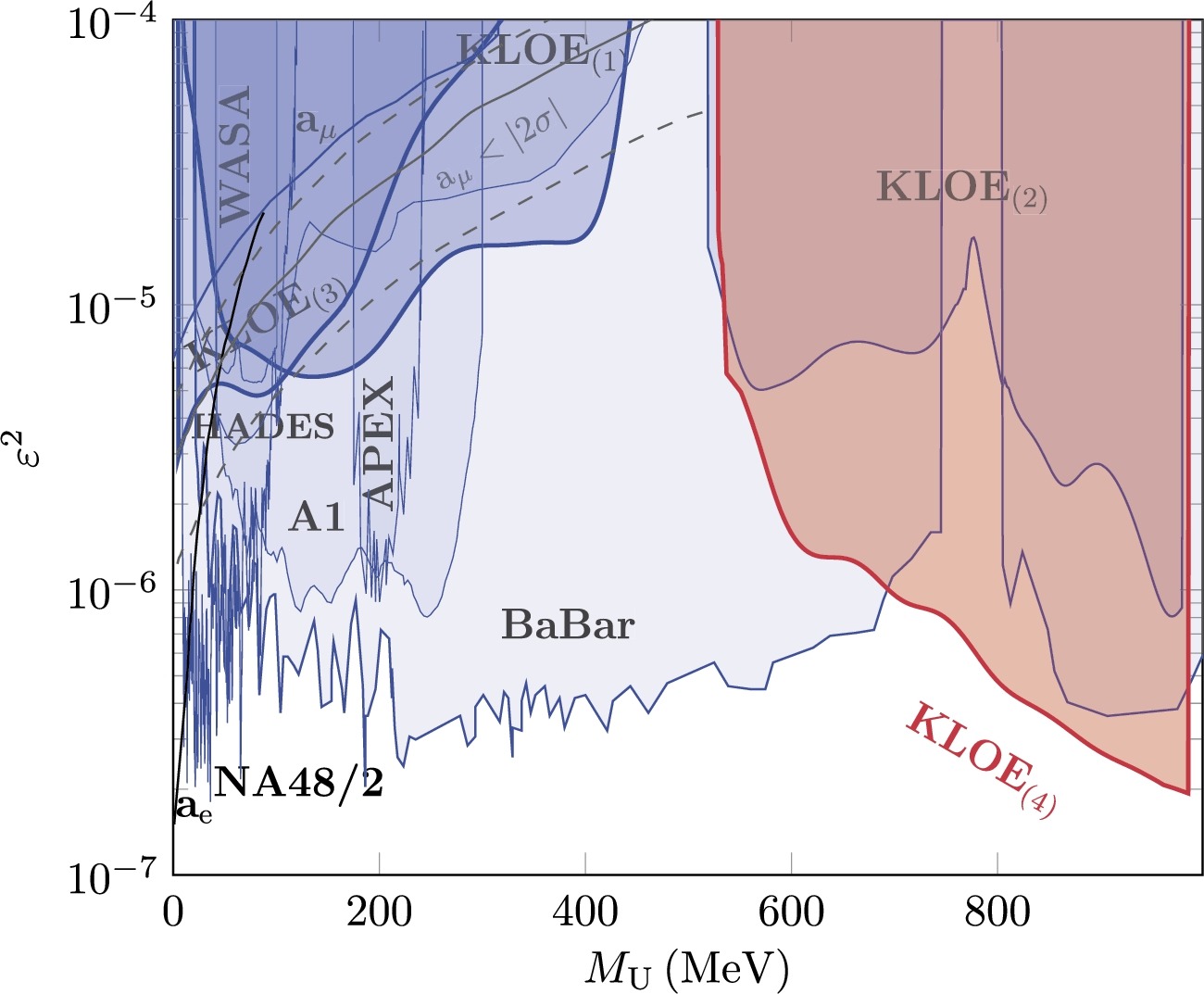}
\end{tabular}
\caption{Monte Carlo simulated fractional backgrounds to the U($\pi^+\pi^-)\gamma$ process after all cuts, normalized to $\pi^+\pi^-\gamma$ (left panel); Comparison of the measured (blue) and simulated (red) two-pion invariant mass with zoomed region of the $\omega -\rho$ interference (central panel); Summary plot of the 90\% confidence levels in $\varepsilon^2$ {\it vs.} M$_{\text{U}}$ plane for the associated production with the $\mu^+\mu^-$, e$^+$e$^-$ and $\pi^+\pi^-$, and $\phi\rightarrow\eta$U with Dalitz U decay (Fig.~2, bottom right). KLOE-2 contours correspond to the processes $\phi\rightarrow\eta$U (KLOE$_1$), $\gamma\,$U(ee) (KLOE$_2$), $\gamma\,$U($\mu\mu)$ (KLOE$_3$) and $\gamma\,$U$(\pi\pi)$ (KLOE$_4$). Confidence levels from other experiments \cite{others,others1,others2,others3,others4,others5} are also shown for comparison. Solid contours represent limits from the electron and muon anomaly \cite{anom}. The grey line shows the U boson parameters that could explain the observed $a_{\mu}$ discrepancy with a $2\sigma$ error band (grey dashed lines) \cite{anom} (right panel).}
\label{fig3}
\end{figure}
Fig. \ref{fig3} (left) presents simulated contributions of backgrounds to the associated U$(\pi^+\pi^-)\,\gamma$ production, after all cuts. 
As seen in the figure, total background contribution does not exceed a few percent.
Quality of the description of the experimental $\pi^+\pi^-$ mass by simulations is presented in Fig.~\ref{fig3}~(central), where data (blue) are shown together with the PHOKHARA Monte Carlo \cite{phokara_gs,phokara_qs1} simulation with the Gounaris-Sakurai pion form factor in the $\omega - \rho$ interference region, kinematically smeared.
As seen from the 90\% confidence level contour plots in Fig.~\ref{fig3} (right), the recent KLOE-2 measurements \cite{u2,u3} constrain the region of the U mass above 600 MeV and $\varepsilon^2$ down to $10^{-7}$, whereas other experiments \cite{others,others1,others2,others3,others4,others5} dominate the region of lower masses. 

\begin{figure}[h]
\begin{center}
\begin{tabular}{c}
\includegraphics[scale=.8]{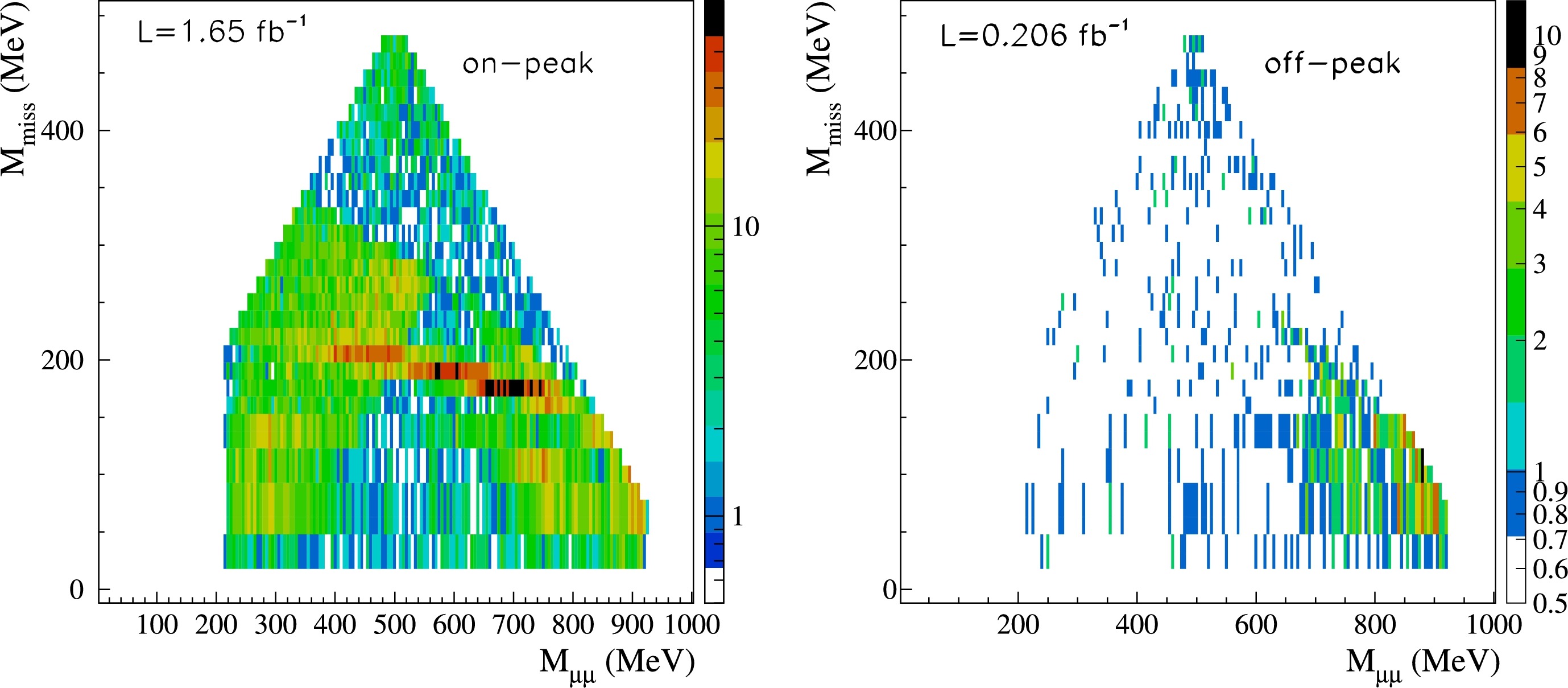} \\
\hspace{-8mm} \includegraphics[scale=.81]{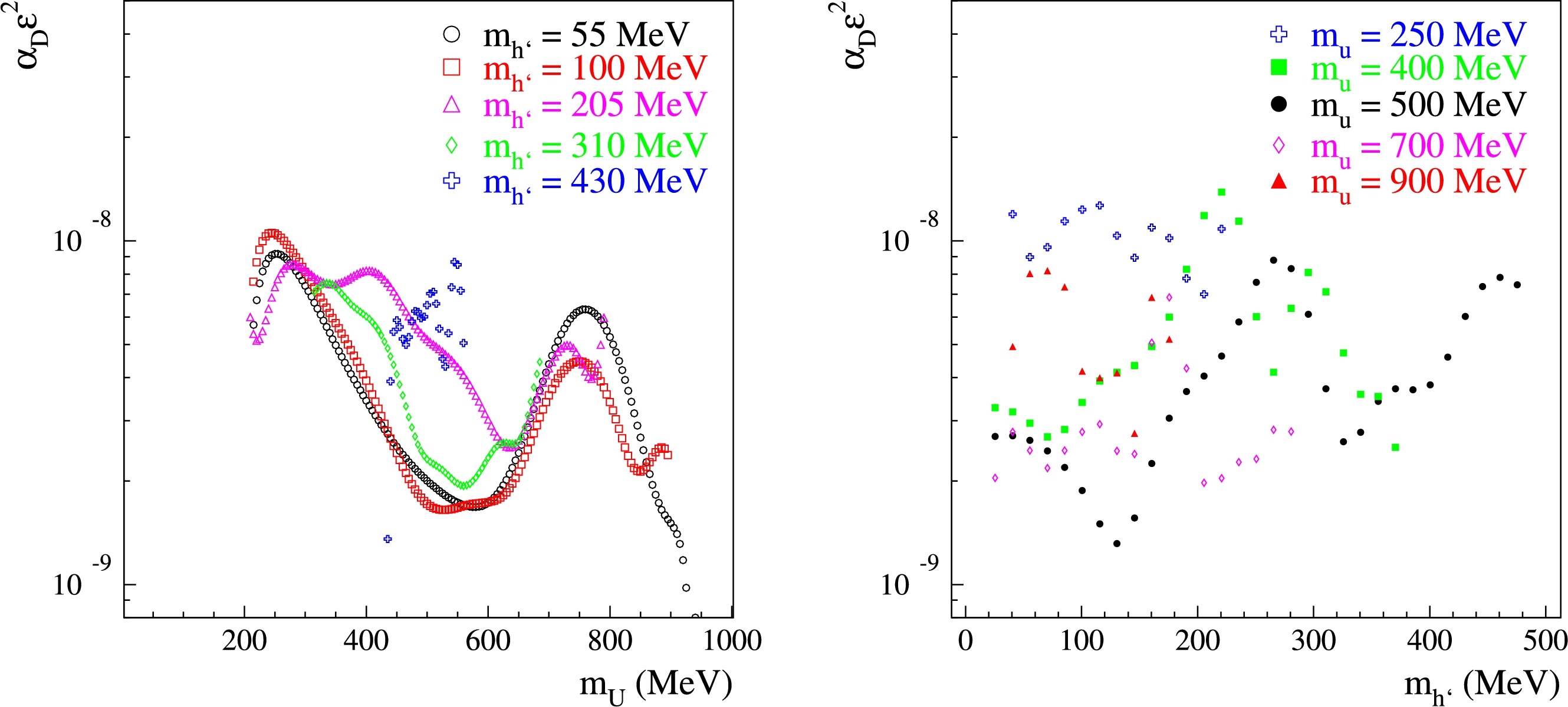}
\end{tabular}
\caption{Background contributions in the missing mass {\it vs.} m$_{\mu\mu}$ plane for the on-peak ($\sqrt{s}=1019$ MeV, top left) and off-peak ($\sqrt{s}=1000$ MeV, top right) subsamples. Backgrounds from the $\phi$ decays are located in the central part (red spots) and the non-resonant backgrounds in the high- and low missing mass regions.
Bottom panels present 90\% confidence levels for $\alpha_D\varepsilon^2$ {\it vs.} $m_U$ or $m_{h^\prime}$ for different mass hypotheses for the Higgs and U bosons, respectively.}
\label{fig4}
\end{center}
\end{figure}

Another interesting process constraining possible existence of the U boson is the Higgstrahlung (cf. Fig.~\ref{fig2}, bottom left), where the photon first couples to U (coupling $\varepsilon$) and then U couples to h$^\prime$ (coupling $\alpha_D$) and then converts into $\mu^+\mu^-$ pair.
Depending on the h$^\prime$ and U mass hypotheses, different search strategies can be followed.
If $m_{h^\prime}>m_U$, the conversion h$^\prime\rightarrow$U can be searched for (strategy taken by the BaBar and Belle Collaborations) and for $m_{h^\prime}<m_U$ the missing mass in the final states is sought (KLOE-2 strategy \cite{higgsstr}). 
In order to better understand backgrounds, two data samples were analysed for this process: the on-peak sample at $\sqrt{s}=1019$ MeV and the off-peak sample, where backgrounds are strongly reduced, at $\sqrt{s}=1000$ MeV (cf. Fig.~\ref{fig4}, top left and right, respectively).
The 90\% confidence limits are plotted as $\alpha_D\varepsilon^2$ {\it vs.} m$_U$ for different masses of h$^\prime$ (Fig.~\ref{fig4}, bottom left) and $\alpha_D\varepsilon^2$ {\it vs.} m$_{h^\prime}$ for different masses of U (Fig.~\ref{fig4}, bottom right).
As seen from these figures, KLOE-2 can reach sensitivity down to $\alpha_D\varepsilon^2\sim 10^{-9}$.

Results from the decay $\phi\rightarrow \eta(3\pi) U$ were published by KLOE-2 in Ref.~\cite{phietau} and show sensitivity to the bottom region of $m_U<500$ MeV (cf. Fig.~\ref{fig3}, right, contour KLOE$_{(1)}$). 

\section{The CP, CPT and Lorentz Symmetries and Quantum Decoherence}

The KLOE-2 has recently obtained results on search for the CP-violating, still unobserved, rare decay K$_S\rightarrow 3\pi^0$ \cite{kl3pi0}.
The K$_S$ mesons are experimentally tagged by the K$_L$ interactions in the calorimeter where they are required to meet timing requirement and deposit more than 100 MeV energy.
Signal candidates are the six-photon events well fitting the vertex and matching energy-momentum balance between the kaon and photons.
No such events were found in a data sample of 77,000 events and KLOE-2 has set new 90\% confidence-level limits on the branching fraction ${\mathcal B}(K_S\rightarrow 3\pi^0)<2.6\times 10^{-8}$ and the amplitude ratio $\eta_{000}=A(K_S\rightarrow 3\pi^0)/A(K_L\rightarrow 3\pi^0)<0.0088$. 

KLOE-2 is also looking for the K$_S\rightarrow \pi^+\pi^-\pi^0$ decays whose amplitude contains also the CP-conserving component.
The measurement is important because both $\eta_{000}$ and $\eta_{+-0}$ contribute to the phase of the mixing CP parameter $\varepsilon$ via Lavoura's relation \cite{lavoura} 
\begin{eqnarray}
\Gamma_{12}=2\pi\sum_f{\mathcal B}(K_S\rightarrow f)[1-|\eta_f|^2-2\Im (\eta_f)], \quad f=3\pi^0,\;\pi^+\pi^-\pi^0. 
\label{eq2}
\end{eqnarray}
KLOE-2 is going to improve limits on the ${\mathcal B}(K_L\rightarrow 3\pi^0)$ and measure ${\mathcal B}(K_S\rightarrow \pi^+\pi^-\pi^0)$ more accurately than existing measurements.

Some Standard Model extensions \cite{kostelecky} allow us to test, using heavy flavour decays, the existence of the CPT-violating and Lorentz-invariance violating term in the Lagrangian, where fermions (quarks) are coupled to the background field via quadruplet of constants $a_\mu$.
Possible violation of the Lorentz invariance entails dependence of the CPT-violating parameter $\delta$ on the sidereal time $\delta=i\,\sin(\varphi)e^{i\varphi}/\Delta m\,\gamma(\Delta a_0-\vec{\beta}\cdot\Delta \vec{a})$, where $\varphi$ stands for the superweak phase depending on $\Delta m= K_L-K_S$ and decay widths, and $\Delta a_{\mu}$ is a difference of couplings to the meson's valence quark and antiquark.
The parameter $\delta$ contributes to intensity spectra of pairs of entangled K$_L$ and K$_S$ from the $\phi$ decay.
\begin{figure}[h]
\begin{tabular}{cc}
\includegraphics[scale=.5]{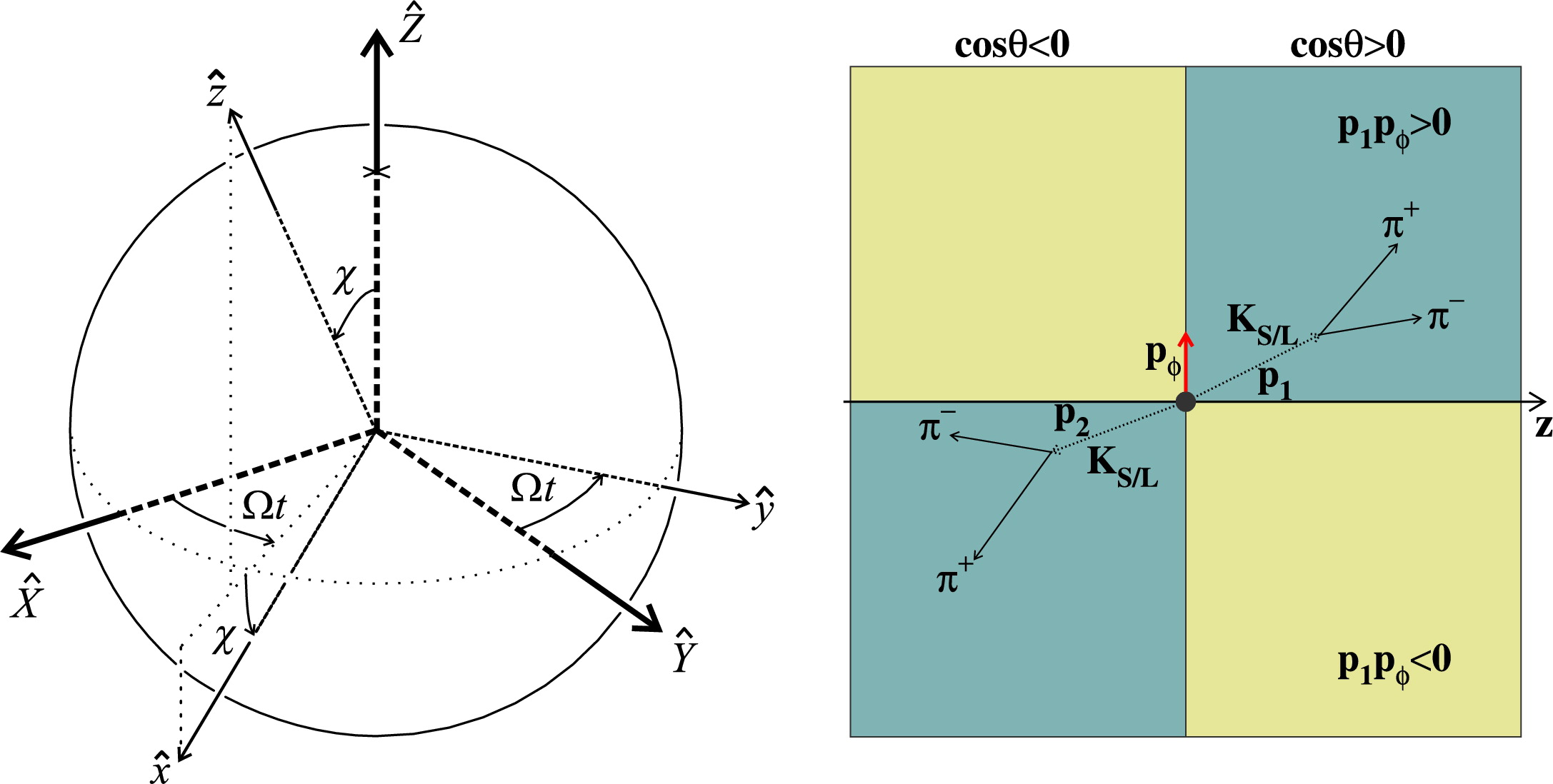} & \hspace{3mm} \includegraphics[scale=.52]{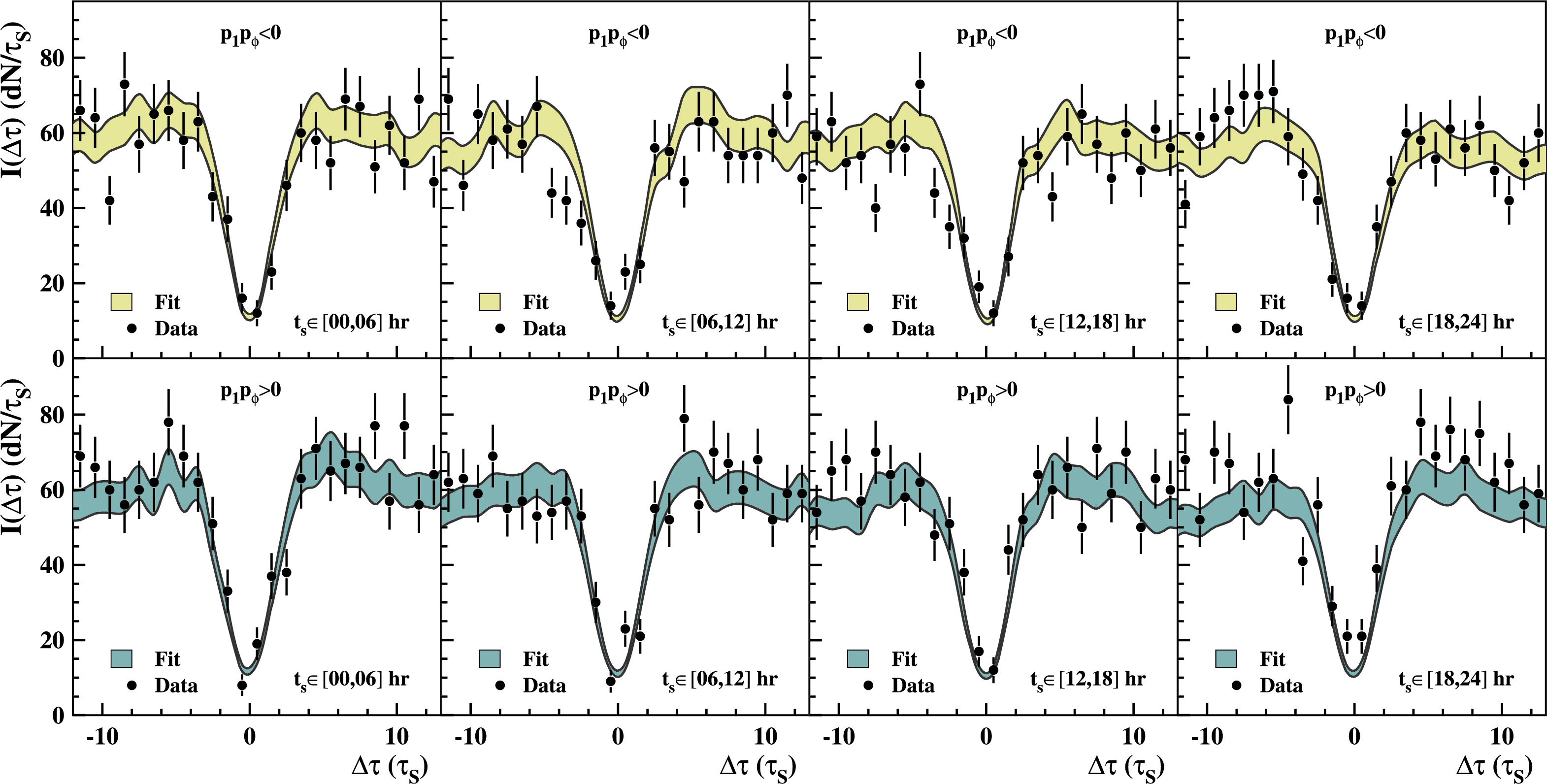}
\end{tabular}
\caption{Definition of the laboratory reference frame parameters for the Earth rotating with angular velocity $\Omega$ and moving with respect to distant stars (left panel); Two orientations of the neutral kaon pairs with respect to $\phi$ (central panel); Spectra of decay times for $K_{L,S}\rightarrow\pi^+\pi^-$ in bins of sidereal time and for two orientations (right panel).}
\label{fig5}
\end{figure}
Fig.~\ref{fig5} presents the KLOE-2 reference frame rotating with the Earth with respect to distant stars (left), four regions of the test variables defining the orientation and sidereal time intervals (center) and the kaon decay-time difference ratios, in units of $\tau_S$, for different orientation and sidereal times (right).
The parameters $\Delta a_{\mu}$ have been determined from the fit \cite{kloe2amu} using decays of both kaons into pairs of charged pions.
Data are consistent with no evidence of the CPT-violating effect but
the experiment is sensitive to $\Delta a_\mu\simeq 10^{-18}$ GeV, close to the Planck's scale $m_K^2/m_p=0.2\times 10^{-19}$ GeV.
This represents by far the best sensitivity achieved experimentally.
Experiments using heavier mesons exhibit accuracy much below that of KLOE-2 \cite{cpt_others,cpt_others1,cpt_others2,cpt_others3}. 

In addition to the sidereal time-dependence studies, future of the CPT research program of KLOE-2 is related to three measurements: (a) search for decoherence in entangled kaon pairs, (b) decay time evolution of the CPT-coupled decay channels, and (c) asymmetries in semi-leptonic decays of K$_L$ and K$_S$.

a) Search for a possible loss of entanglement due to possible quantum background effect, as e.g. gravity below the nanometer scale, necessarily entails evolution of the pure into the mixed state and violates CPT.
First results by KLOE \cite{decoherence,decoherence1} demonstrated that breakdown of the Furry hypothesis can be tested at the level of $10^{-7}$, dissipative decoherence parameters can be determined with accuracy of $10^{-17}-10^{-19}$ GeV, and admixture of the wrong-statistics states to the entangled two-kaon states down to the level of $10^{-4}$.  
An improvement by KLOE-2 is expected due to higher luminosity and four times better resolution in decay time.

b) As noticed in Ref.~\cite{bernabeu}, the two-kaon observable
\begin{eqnarray}
R_{2,4}(\Delta t)=\frac{I(K_{e^{\pm}3},3\pi^0;\Delta t)}{I(2\pi,K_{e^{\pm}3};\Delta t)}\frac{{\mathcal B}(K_L\rightarrow 3\pi^0)}{{\mathcal B}(K_S\rightarrow 2\pi)}\frac{\Gamma_L}{\Gamma_S}
\label{eq3}
\end{eqnarray} 
depends on $\Re(\delta)$. 
In particular, in the asymptotic regime
\begin{eqnarray}
R_{2,4}(\Delta t) \sim 1\mp \Re(\delta),\quad\quad\quad\mbox{for}\quad \Delta t\gg \tau_S.
\label{eq4}
\end{eqnarray} 
Statistical sensitivity of this observable is $3- 1.5\times 10^{-3}$ for a KLOE-2 luminosity $5-20$ fb$^{-1}$. 
KLOE-2 will provide a new measurement of $\Re(\delta)$ using different method than CPLEAR \cite{cplear}.

c) The last ongoing component of the KLOE-2 CPT program is the CPT-invariance test using  asymmetries of the semi-leptonic decays 
\begin{eqnarray}
A_{S,L} & = & \frac{\Gamma(K_{S,L}\rightarrow\pi^-e^+\nu_e)-\Gamma(K_{S,L}\rightarrow\pi^+e^-\bar\nu_e)}{\Gamma(K_{S,L}\rightarrow\pi^-e^+\nu_e)+\Gamma(K_{S,L}\rightarrow\pi^+e^-\bar\nu_e)} \nonumber \\
        & = & 2[\Re(\varepsilon)\pm\Re(\delta)]-2[\Re(y)\mp\Re(x_-)],
\label{eq5}
\end{eqnarray}
where $y$ parametrizes CPT violation assuming $\Delta S=\Delta Q$ rule and $x_-$ is a small term describing a possible violation of the $\Delta S=\Delta Q$ rule, experimentally limited to $-0.002\pm 0.006$ \cite{yang}.
The difference $A_S-A_L=4[\Re(\delta)-\Re(x_-)]$ does not depend on $y$.
KLOE-2 aims to improve the current accuracy $\sigma(A_S)=0.01$ \cite{kloe_as} which is now worse than $\sigma(A_L)=0.7\times 10^{-4}$.  

\section{Determination of Transition Form Factors of $\phi$ to Pseudoscalar Mesons}

Precise determination of the transition form factors (TFF) of mesons is crucial for understanding the low-energy structure of hadrons.
In particular, TFFs play an important role in the determination of the light-by-light contribution to the anomalous magnetic moment of the muon \cite{hye,jaegerlehner}. 
The interest is reinforced by failure to describe the data on $\omega\rightarrow\pi^0\mu^+\mu^-$ decay by the Vector Meson Dominance model \cite{omega_others,omega_others1}.
\begin{figure}[h]
\begin{tabular}{cc}
\includegraphics[scale=.55]{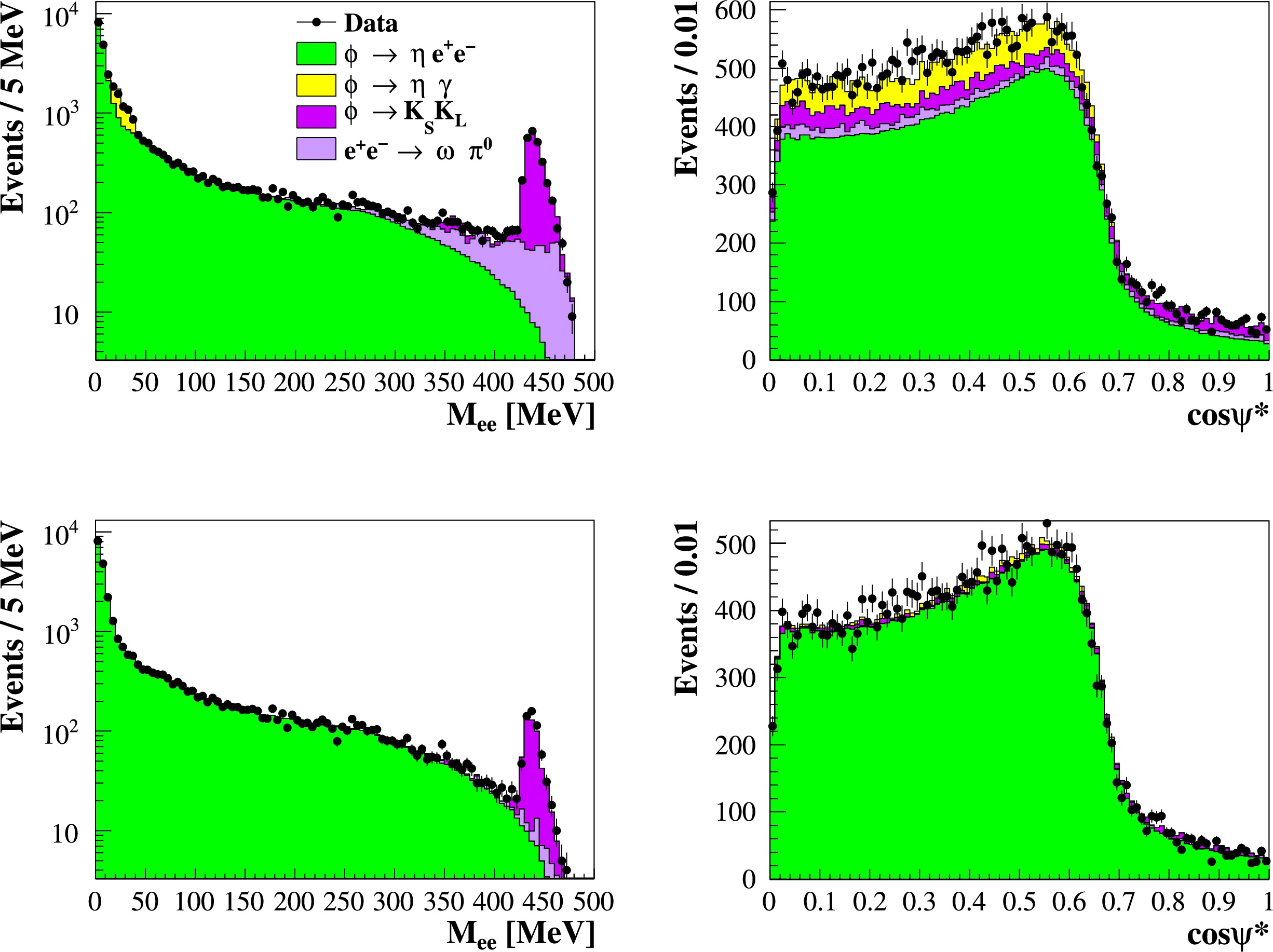} & \hspace{0mm} \includegraphics[scale=.9]{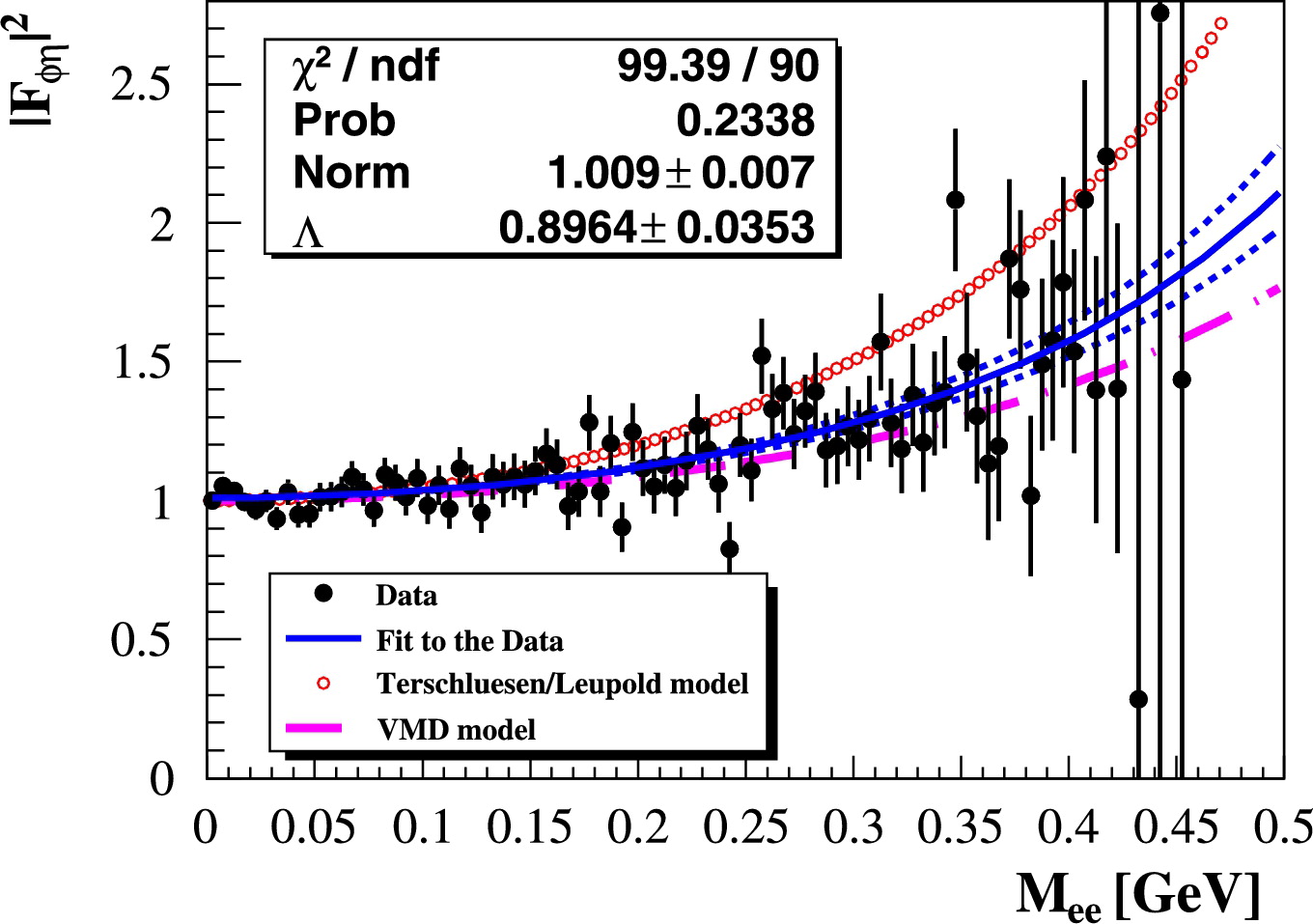}
\end{tabular}
\caption{Left panels: plots of the signal and background contributions to $\phi\rightarrow\eta(3\pi^0)e^+e^-$ in the $e^+e^-$ mass and the $\eta e^+$ angle $\psi^\ast$ before (top) and after (bottom) analysis cuts; right panel: the form factor of the $\phi\eta$ vertex where blue line is the fit while the other two lines are theoretical expectations.}
\label{fig6}
\end{figure}

Recently, KLOE-2 has published new results on TFFs and branching fractions for decays of $\phi$ into pseudoscalar mesons: $\phi\rightarrow\eta(3\pi^0)e^+e^-$ \cite{tff1} and $\phi\rightarrow\pi^0 e^+e^-$ \cite{tff2}. 

Fig.~\ref{fig6} (right and central) presents distributions of the $e^+e^-$ mass and the angle between $e^+$ and $\eta$ for the signal and backgrounds before and after cuts, for the $\phi\rightarrow\eta(3\pi^0)e^+e^-$.
The TFF dependence on $m_{ee}$ is presented (right panel) with fits.
The slope parameter for this TFF was found to be $b_{\eta\phi}(q^2=0)=(1.28\pm 0.10^{+0.09}_{-0.08})$ GeV$^{-2}$ and the branching fraction ${\mathcal B}(\phi\rightarrow \eta e^+e^-)=(1.075\pm 0.007\pm 0.038)\times 10^{-4}$ (the first error is statistical and the second is systematic). 

\begin{figure}[h]
\begin{tabular}{cc}
\includegraphics[scale=.45]{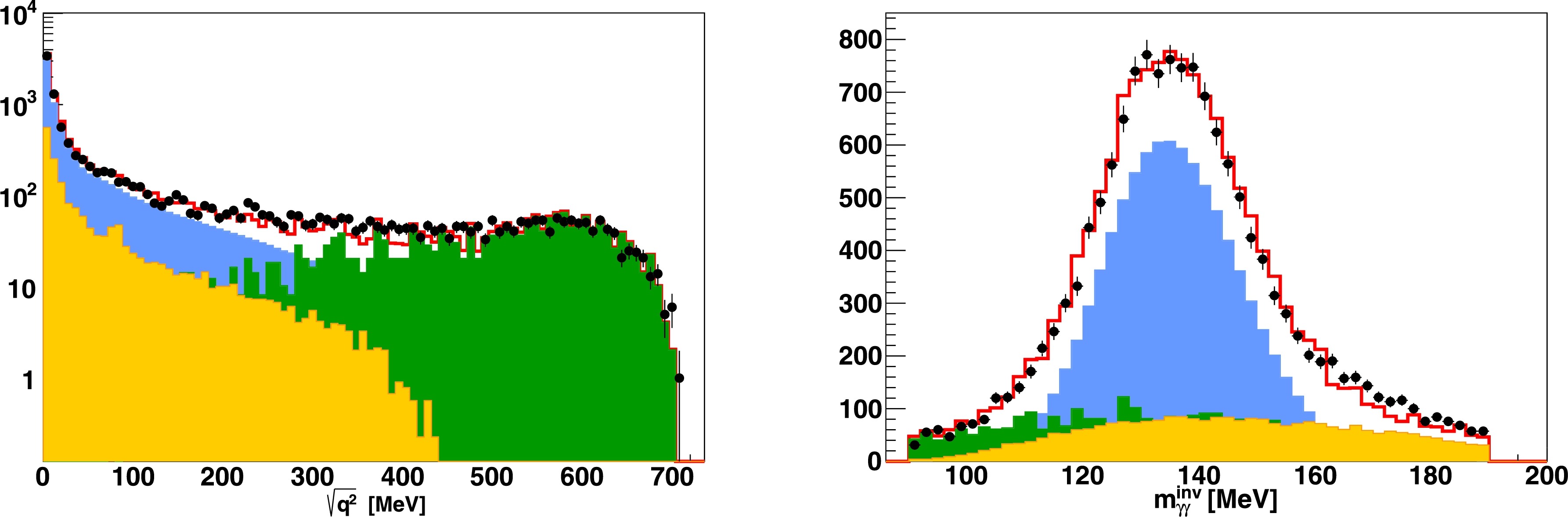} & \hspace{0mm} \includegraphics[scale=.7]{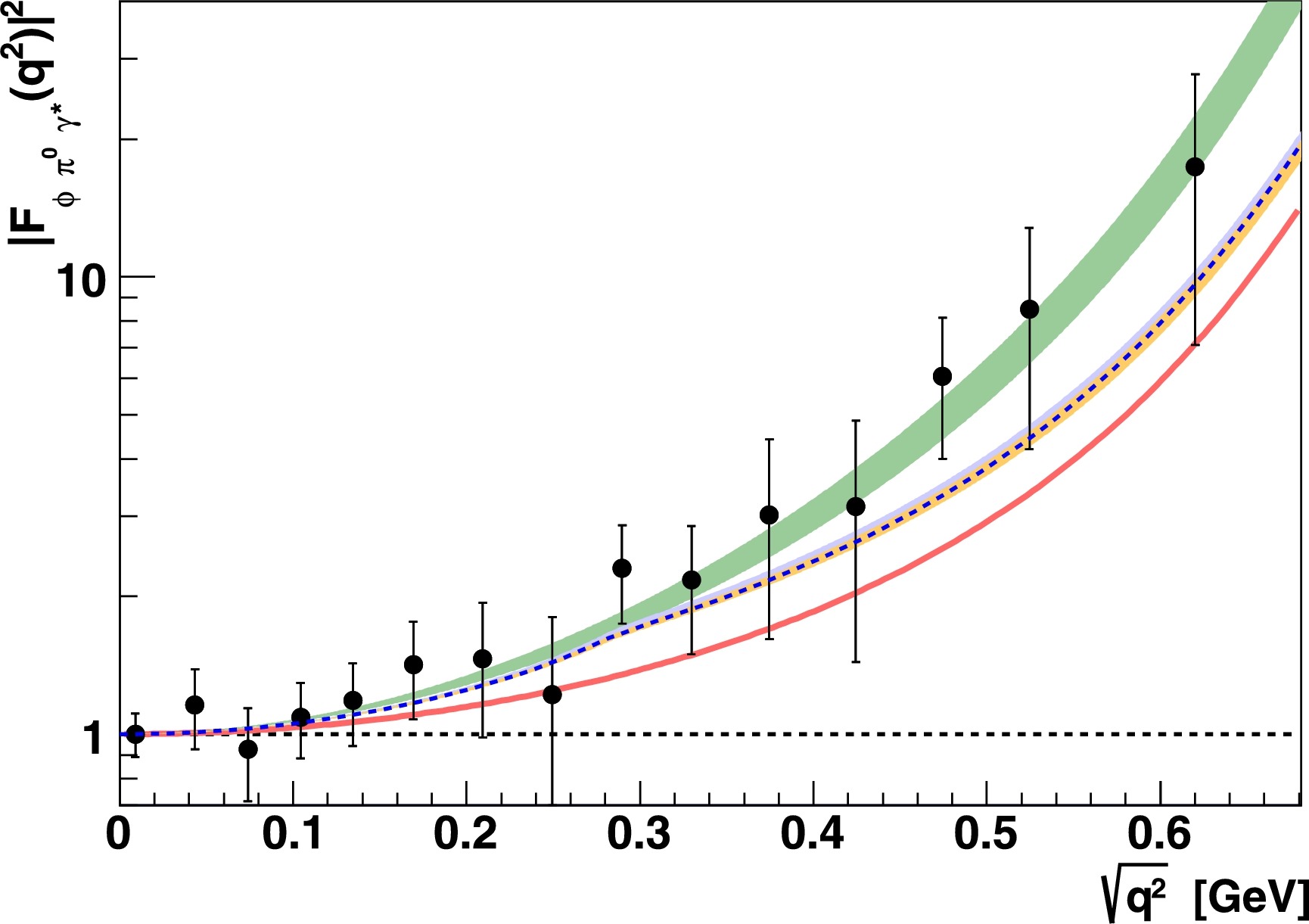}
\end{tabular}
\caption{Left and central panels: plots of the signal and background contributions to $\phi\rightarrow\pi^0e^+e^-$ in variables $\sqrt{q^2}$ and $m_{\gamma\gamma}$; right panel: the form factor of the $\phi\rightarrow\pi^0\gamma^\ast(e^+e^-)$ vertex {\it vs.} $\sqrt{q^2}$ with theoretical models superimposed \cite{fits,fits1,fits2,fits3}: Schneider (dotted), Ivashyn (green), Danilkin (orange), Landsberg (red).}
\label{fig7}
\end{figure}
Similar results for the  $\phi\rightarrow\pi^0e^+e^-$ decay are shown in Fig.~\ref{fig7} where distributions of the $\sqrt{q^2}$ and $m_{\gamma\gamma}$ for the signal and backgrounds are given in the left and central panels.
The right panel presents $|F_{\phi\pi^0\gamma^\ast}(q^2)|^2$ with fits \cite{fits} superimposed.
As seen from Fig.~\ref{fig7}, the Ivashyn model is favoured by the data.
The slope parameter was found to be $b_{\eta\phi}(q^2=0)=(2.02\pm 0.11)$ GeV$^{-2}$ (combined errors) and the branching fraction ${\mathcal B}(\phi\rightarrow \pi^0 e^+e^-)=(1.35\pm 0.05^{+0.05}_{-0.10})\times 10^{-5}$ (the first error is statistical and the second is systematic).

\section{Conclusions}
Using enhanced luminosity of DA$\Phi$NE and with new detectors, KLOE-2 is in the middle of data taking and is aiming to take more than 5 fb$^{-1}$ of data.
The Collaboration searches for new physics around 1 GeV, tests the CP, CPT and Lorentz invariance, and performs precision measurements in hadronic low-energy physics.
Significant progress will be made in all results concerning tests of CPT, CP, quantum mechanics and low-energy structure of mesons.

\acknowledgments
This work was supported by the Polish National Science Center grant nr 2013/08/M/ST2/00323 and EU Hadron Physics Project under contract number RII3-CT-2004-506078; by the EC under the 7$^{th}$ Framework Programme through the "Research Infrastructures" action of the "Capacities" Programme, Call: FP7-INFRASTRUCTURES-2008-1, Grant Agreement No. 227431.

\end{document}